\documentclass[10pt,a4paper,final]{article}
\usepackage[utf8]{inputenc}
\usepackage[english]{babel}
\usepackage{amsmath}
\usepackage{amsfonts}
\usepackage{amssymb}
\usepackage{subcaption} 
\usepackage{graphicx}
\usepackage[left=2cm,right=2cm,top=2cm,bottom=2cm]{geometry}
\usepackage{color,soul} 
\usepackage{cite} 
\usepackage{authblk} 
\usepackage{adjustbox} 
\author[1*]{Juan I. Rodríguez}
\author[2]{Ulises  A.  Vergara-Beltrán}
\affil[1]{Centro de Investigación en Ciencia Aplicada y Tecnología Avanzada, Unidad Querétaro, Instituto Politécnico Nacional, Cerro Blanco 141 Col. Colinas del Cimatario,  Querétaro CP 76090, México.}
\affil[2]{Escuela Superior de F\'{i}sica y Matemáticas, Instituto Politécnico Nacional, Edificio 9, Zacatenco. Col. San Pedro Zacatenco, Ciudad de México, C.P. 07738, México}

\affil[*]{Author to whom correspondence should be addressed: jirodriguezh@ipn.mx}
\date{}                     
\title{A physics-inspired evolutionary machine learning method: from the Schrödinger equation to an orbital-free-DFT kinetic energy functional}
\begin{document}
\begin{titlepage}
\maketitle
\begin{abstract}
We introduce a machine learning (ML) supervised model function that is inspired by the variational principle of physics.  This ML hypothesis evolutionary method, termed ML-$\Omega$, allows us to go from data to differential equation(s) underlying the physical (chemical, engineering, etc.) phenomena the data are derived from. The fundamental equations of physics can be derived from this ML-$\Omega$ evolutionary method when provided the proper training data. By training the ML-$\Omega$ model function with only three hydrogen-like atom energies, the method can find Schrödinger’s exact functional and, from it, Schrödinger’s fundamental equation. Then, in the field of density functional theory (DFT), when the model function is trained with the energies from the known Thomas-Fermi (TF) formula $E=-0.7687Z^\frac{7}{3}$, it correctly finds the exact TF functional. 
Finally, the method is applied to find a local orbital-free (OF) functional expression of the independent electron kinetic energy functional $T_s$ based on the $\gamma TF\lambda vW$ model.
By considering the theoretical energies of only 5 atoms (He, Be, Ne, Mg, Ar) as the training set, 
the evolutionary ML-$\Omega$ method finds an ML-$\Omega$-OF-DFT local $T_s$ functional  
($\gamma TF \lambda vW(0.964,1/4)$) that outperforms all the OF-DFT functionals of a representative group. 
Moreover, our ML$\Omega$-OF functional overcomes the LDA’s and some local GGA-DFT’s functionals' difficulty to describe the stretched bond region at the correct spin configuration of diatomic molecules. Although our evolutionary ML-$\Omega$ model function can work without an explicit prior-form functional, by using the techniques of symbolic regression, in this work we exploit prior-form functional expressions to make the training process faster in the example problems presented here.  

\end{abstract}

\end{titlepage}

The laws of nature are subject to certain types of optimization processes. In classical mechanics, this was given a rigorous mathematical formalism as Hamilton’s principle of least-action \cite{lanczos2012variational,yourgrau,goldstein2011classical}. Accordingly, equations of physics can be deduced from this principle with an “appropriate” Lagrangian. \cite{lanczos2012variational} In his original publication, Schrödinger deduced his eponymous equation for the hydrogen atom from a variational problem.\cite{schrodingjsr19262,schrodinger2003collected}. Similarly, density functional theory (DFT) is based on a variational problem formulated as the second Hohenberg-Kohn theorem \cite{hohenberg1964inhomogeneous,dreizler2012density,ParrYang1995}.  All of these variational processes represent the minimization of a functional subject to some restrictions, a typical stationary problem in the Calculus of Variations \cite{gelfand2000calculus,Elsgole}.

In this work, we introduce a machine-learned, ML-$\Omega$, hypothesis function for use in supervised learning which is defined as: 

\begin{equation}
\label{equ: ML_model_function}
\Omega(\vec{x}; \vec{W}) \equiv \underset{\phi}{\mathbf{min}} \; J_{\alpha}[\phi;\vec{x},\vec{W}] = \underset{\phi}{\mathbf{min}} \int_D F_{\alpha}(\phi,\nabla \phi, \nabla \nabla \phi, \vec{r}; \vec{x}, \vec{W})d\vec{r}, 
\end{equation}
the minimization in Eq. (\ref{equ: ML_model_function}) is subject to $n_r$ restrictions on the function $\phi:D \subseteq \mathfrak{R}^d \rightarrow  \mathfrak{R}$,

\begin{equation}
\label{equ: ML_restrictions}
 \mathfrak{F}_i (\phi) = 0 ; i=1,2,..., n_r  .
\end{equation}
$\vec{x} \in  \mathfrak{R}^d$ and $\vec{W} \in \mathfrak{R}^m$ are the feature and weight vectors, respectively; $J$ is a functional of the scalar field $\phi$; $\nabla \phi$ is the gradient of $\phi$ ; $\nabla \nabla  \phi$ represents the partial derivatives of $\phi$ of superior order; while $\alpha$ represents a set of hyper-parameters. The training data is of the type $(\vec{x_i }, t_i)$ , $\vec{x}_i \in \mathfrak{R}^d$ , $t_i \in \mathfrak{R} ,i=1, \cdots , K$. The minimization in Eq. (\ref{equ: ML_model_function}) is performed   
with the meta-heuristic differential evolution (DE) global optimization method recently introduced by us, which is described in detail in reference \cite{vergara2023efficient}. 

The loss function,

\begin{equation}
\label{equ:Loss_function}
\Delta (\vec{W}) = \frac{1}{K} \sum_{i=1}^K L(\Omega(\vec{x}_i;\vec{W})- t_i) + \Theta(\vec{W}),
\end{equation}
is minimized using our evolutionary DE method in all cases considered in this work. In Eq.(\ref{equ:Loss_function}), $L$ represents a norm and $\Theta$ a regularization function. Once the model function is trained by minimizing the loss function Eq. (\ref{equ:Loss_function}), the target value for a new feature vector $\vec{x}_{\mu}$ that is not in the training set is predicted,

\begin{equation}
\label{equ:target_value_mu}
t_{\mu} = \Omega(\vec{x}_{\mu}; \vec{W}).
\end{equation}

An interesting and powerful feature of the model function is that the minimization inherent to it (Eq. (\ref{equ: ML_model_function})) is equivalent to solving the corresponding Euler-Lagrange equation for $\phi $. Thus, the ML model function allows one to go directly from training data to a differential equation, which itself potentially describes the underlying physics, chemistry, or engineering phenomena.

Before applying this method to the open problem of finding approximations to the DFT energy functional, let us first apply it to two exact fundamental equations in physics as a proof of concept. In his original publication, as remarked earlier, Schrödinger applied a variational process \cite{schrodingjsr19262,schrodinger2003collected}. In an effort to reproduce the experimentally observed hydrogen atomic spectrum from a differential wave-like equation, Schrödinger proposed minimizing the following functional (in this work atomic units (a.u.) are used throughout)\footnote{Schrödinger actually focused on the particular case of the hydrogen atom: $\mu = Z = 1$}:

\begin{equation}
\label{equ: Schrodinger_functional}
J[\Psi] = \frac{1}{2 \mu}\int_{\mathfrak{R}^3} |\nabla \Psi|^2 d\vec{r} -Z\int_{\mathfrak{R}^3} \frac{\Psi^2}{r} d\vec{r}
\end{equation}
subject to the restriction,

\begin{equation}
\label{equ: Schrodinger_restriction}
\int_{\mathfrak{R}^3} \Psi^2 d\vec{r} = 1,
\end{equation}
which resulted in the equation for the hydrogen-like atom,

\begin{equation}
\label{equ: Schrodinger_equation}
\left[-\frac{1}{2 \mu} \nabla^2 + V(\vec{r}) \right] \Psi = E \Psi,
\end{equation}
where $V(\vec{r})= -\frac{Z}{r}$; $Z$ is the atomic number, $\mu$ is the reduced mass, and the energy E is the Lagrange multiplier that fulfils the restriction Eq.(\ref{equ: Schrodinger_restriction}).
As a proof of concept, let us suppose that the functional Eq. (\ref{equ: Schrodinger_functional}) (Equation (\ref{equ: Schrodinger_equation})) is not known, and that the experimental ground state energies of the hydrogen-like atoms are available.  Then let us suppose that our model function $\Omega$ is of the form:

\begin{equation}
\label{equ: like_Schrodinger_functional}
\Omega(\vec{x},\vec{W}) = \underset{\phi}{\mathbf{min}} \; Y [\phi;\vec{x},\vec{W}] = \underset{\phi}{\mathbf{min}}\; \frac{1}{2\mu} \int_{\mathfrak{R}^3} |\nabla \phi|^{w_1} d\vec{r} -Z \int_{\mathfrak{R}^3} \frac{\phi^{w_2}}{r}d\vec{r},  
\end{equation}
where the minimization is subject to the restriction, 

\begin{equation}
\label{equ: like_Schrodinger_restriction}
\int_{\mathfrak{R}^3} \phi^2 d\vec{r} = 1.
\end{equation}
In this case, the feature vector has two entries, the atomic number and the reduced mass, $\vec{x_i} = (Z_i,\mu_i)$; the weight vector $\vec{W} = (w_1,w_2)$ has also two components. As a training set, we consider the ground state energy of the hydrogen-like atoms for Z=1-3 ($H$,$He^{+}$, $Li^{2+}$) as the target values.\footnote{Instead of using the experimental values, the energy values are obtained from the known formula,$E_{1S} = -\frac{Z^2\mu}{2}$.}  The Cartesian $L_2$ norm is considered in the loss function with no regularization for all cases studied in
this work. 
The DE minimization method used in the ML-$\Omega$ model function relies on an expansion of $\phi$ on a finite basis set, 

\begin{equation}
\label{equ: bs_expansion}
\phi(\vec{r}) = \sum_{i=1}^M c_i \chi_i(\vec{r}).
\end{equation}
 Lagrange multipliers are not used to satisfy restriction Eq.(\ref{equ: like_Schrodinger_restriction}) but the following restrictions for the coefficients and basis functions,
$c_i = c_i/ \sqrt{ C_r } \: ; \: C_r = \sum_{i=1}^M \sum_{j=1}^M  c_i c_j \int_{\mathfrak{R}^3} \chi_i(\vec{r}) \chi_j(\vec{r})  d\vec{r}$.
Gaussian basis functions were used $\chi(\vec{r}) = x^{l}y^{m}z^{n} \exp(-\alpha r^2)$ 
with $l=m=n=0$ and $\alpha = 1.6^\zeta \;; \zeta = -13,-12,\cdots,38$. 
Both integrals from the ML-$\Omega$ functional Eq.(\ref{equ: like_Schrodinger_functional}) were computed numerically using a Gauss-Legendre quadrature. 

From Table \ref{tab: w_values_S_and_TF} we can see that the training process converged to the exact answer up to the 8th and 4th decimal places for $w_1$ and $w_2$, respectively, in the Schrödinger functional. Notice that although the proposed functional form in the ML-$\Omega$ model function (Eq. (\ref{equ: ML_model_function})) is quite similar to the exact functional, the weights $w_i$ could in principle take an infinite number of different values, which in turn would produce an infinite number of different functionals. However, the training process finds the correct exact Schrödinger's functional, which can be analytically minimized to obtain the corresponding Euler-Lagrange equation that, in this case, is the Schrödinger equation. Thus, the ML-$\Omega$ method proposed here finds the Schrödinger equation from the "experimental" data available.

In the second proof-of-concept problem, in the realm of density functional theory, we perform a similar process to find the Thomas-Fermi (TF) functional \cite{thomas1927calculation,fermi1927statistical}. The proposed ML model function considered in this case is (now $\phi$ "represents" the electron density $\rho$ instead of the wave function),

\begin{equation}
\label{equ: TF_like_functional}
\Omega(\vec{x},\vec{W}) = \underset{\phi}{\mathbf{min}} \; Y [\phi;Z,\vec{W}] = \underset{\phi}{\mathbf{min}} \;\int_{\mathfrak{R}^3} w_1 \phi^{w_2}(\vec{r}) d\vec{r} - Z\int_{\mathfrak{R}^3}\frac{\phi(\vec{r})}{r} d\vec{r} +\frac{1}{2} \iint_{\mathfrak{R}^6} \frac{\phi(\vec{r})\phi(\vec{r'})d\vec{r} d \vec{r'}}{|\vec{r}- \vec{r'}|}. 
\end{equation}
The minimization in Eq.(\ref{equ: TF_like_functional}) must be performed considering the following restrictions on $\phi$:

\begin{equation}
\label{equ: TF_restriction_normlaization}
\int_{\mathfrak{R}^3} \phi(\vec{r})d\vec{r} = N,
\end{equation}

\begin{equation}
\label{equ: TF_restriction_positivity}
\phi(\vec{r}) \geq 0 \; \forall \vec{r} \in \mathfrak{R}^3,
\end{equation}
which are satisfied by imposing the following conditions on the expansion coefficients and basis functions: 
$\sum_j^n c_j = N$, $0 \leq c_j \leq N$, $\chi_j(\vec{r}) \geq 0 \quad \forall \vec{r} \in \mathbf{R}^3$,
$\int \chi_j(\vec{r}) d\vec{r} = 1, \forall j$. 
In this case, the feature vector is a scalar (Z) and the target is the TF energy. So the training data is of the form $(Z_i,E_{TF}(Z_i))$,
which is again considered only for three first atoms ($H$,$He$,$Li$). Energy values are obtained from the known formula for the so-called
TF-Dirac model for neutral atoms $E_{TFD}(Z) = -0.7687Z^{\frac{7}{3}}$ \cite{ParrYang1995,march2016self}.
Atom centered squared Gaussian basis functions were used for this case. The external potential and Coulomb integrals were
performed analytically; the rest of the integrals were performed numerically.
From Table \ref{tab: w_values_S_and_TF} we can see that the training process effectively finds the exact values for the weights up to the third and fourth decimal places, respectively. 
It is important to keep in mind that the DE minimization method used in the ML-$\Omega$ model function relies on the expansion of $\phi$ on a finite basis set ( Eq. (\ref{equ: bs_expansion}))\cite{vergara2023efficient}. A steady convergence of the weights to the exact TF functional coefficients and exponents is obtained as the basis set used in ML-$\Omega$ (Eq. (\ref{equ: TF_like_functional}) ) is increased.

\begin{table}

\centering
\caption{Exact parameters and ML-$\Omega$ weights.}
\label{tab: w_values_S_and_TF}
\begin{tabular}{c|c|c}
	& $w_1$ & $w_2 $ \\
\hline
\hline	
Schrödinger exact values & 2 & 2 \\
ML-$\Omega$ functional  & 2.00000 & 2.00003 \\
\hline
\hline
Thomas-Fermi exact values & 2.87123 & 5/3 \\
ML--$\Omega$-TF functional & 2.87085 & 1.66663 \\
\hline
\hline
 ML--$\Omega$-OF-DFT 	 & 0.964 & 0.250  \\

\end{tabular}
\end{table}

Recently, machine learning has been used to find orbital-free (OF) expressions of the independent electron kinetic energy functional $T_s$ using standard ML hypothesis functions like kernel ridge regression, \cite{kulik2022roadmap,li2016understanding,meyer2020machine}  and convolutional neural networks \cite{meyer2020machine,kulik2022roadmap}. Although these works have reported promising results, they still suffer from a lack of universality. A new training process has to be performed for each new type of molecule. The training process, however, implies solving the electronic structure problem several times using the standard CPU-high-cost methods, which is actually what one hopes to avoid by using an OF-DFT functional from the outset. Here we apply our ML-$\Omega$ method to find an OF expression for $T_s$ using a simple idea. 
 In the Kohn-Sham DFT (KS-DFT) scheme, the electron gas is modelled by an independent electron virtual system whose energy is,

\begin{equation}
\label{equ: Kohn-Sham_functional}
E[\rho] = T_s[\rho]  + E_{xc}[\rho] + \iint_{\mathfrak{R}^6} \frac{\rho(\vec{r})\rho(\vec{r'})d\vec{r} d \vec{r'}}{|\vec{r}- \vec{r'}|} + \int_{\mathfrak{R}^3} v_{ext}(\vec{r})\rho(\vec{r})d\vec{r},
\end{equation}
where  $E_{xc}[\rho]$ is the  exchange-correlation (xc) energy and $v_{ext}$ is the external potential.  $T_s[\rho]$  is actually  
an explicit functional of the orbitals and an implicit functional of $\rho$, 

\begin{equation}
\label{equ: T_s_Kinetic_energy_orbitals}
T_s[\Psi_1,\cdots,\Psi_N] = \sum_{i=1}^N \int_{\mathfrak{R}^3} \Psi^*_i(\vec{r})\left(-\frac{\nabla^2}{2}\right)\Psi_i(\vec{r})d \vec{r},
\end{equation}

\begin{equation}
\rho(\vec{r}) = \sum_{i=1}^N f_i|\Psi_i(\vec{r})|;
\end{equation}
where $\Psi_i(\vec{r})$ and $f_i$ ; $i=1,\cdots,N$ are the orbitals and occupation numbers, respectively \cite{kohn1965self,ParrYang1995,dreizler2012density}. 
As a simple model example, we propose here to optimize the parameters of the TF-Von Weizsaker model functional ($\gamma TF\lambda vW$),\cite{ParrYang1995,dreizler2012density} so that the ML-$\Omega$ model function is in this case:

\begin{equation}
\label{equ: Ts_OFDFT_ML}
\Omega(\vec{x},\vec{W}) = \underset{\phi}{\mathbf{min}} \; Y [\phi;\vec{x},\vec{W}] = \underset{\phi}{\mathbf{min}} \; T_s[\phi;\vec{W}] +E_{xc}[\phi] + \int_{\mathfrak{R}^3}v_{ext}(\vec{r};\vec{x})\phi(\vec{r}) d\vec{r} +\frac{1}{2} \iint_{\mathfrak{R}^6} \frac{\phi(\vec{r})\phi(\vec{r'})d\vec{r} d \vec{r'}}{|\vec{r}- \vec{r'}|};
\end{equation}

\begin{equation}
\label{equ: Ts_OFDFT}
T_s[\phi;\vec{W}] = \int_{\mathfrak{R}^3} \left[ w_1c_{TF} \phi^{\frac{5}{3}}(\vec{r}) + w_2\frac{1}{8}\frac{|\nabla \phi (\vec{r})|^{2}}{\phi(\vec{r})}\right] d\vec{r},
\end{equation}
where $c_{TF} = 2.87123 $ is the Thomas-Fermi constant. 
  
\begin{figure}[ht]
\begin{subfigure}{1\linewidth}
\centering
\includegraphics[width=1\linewidth]{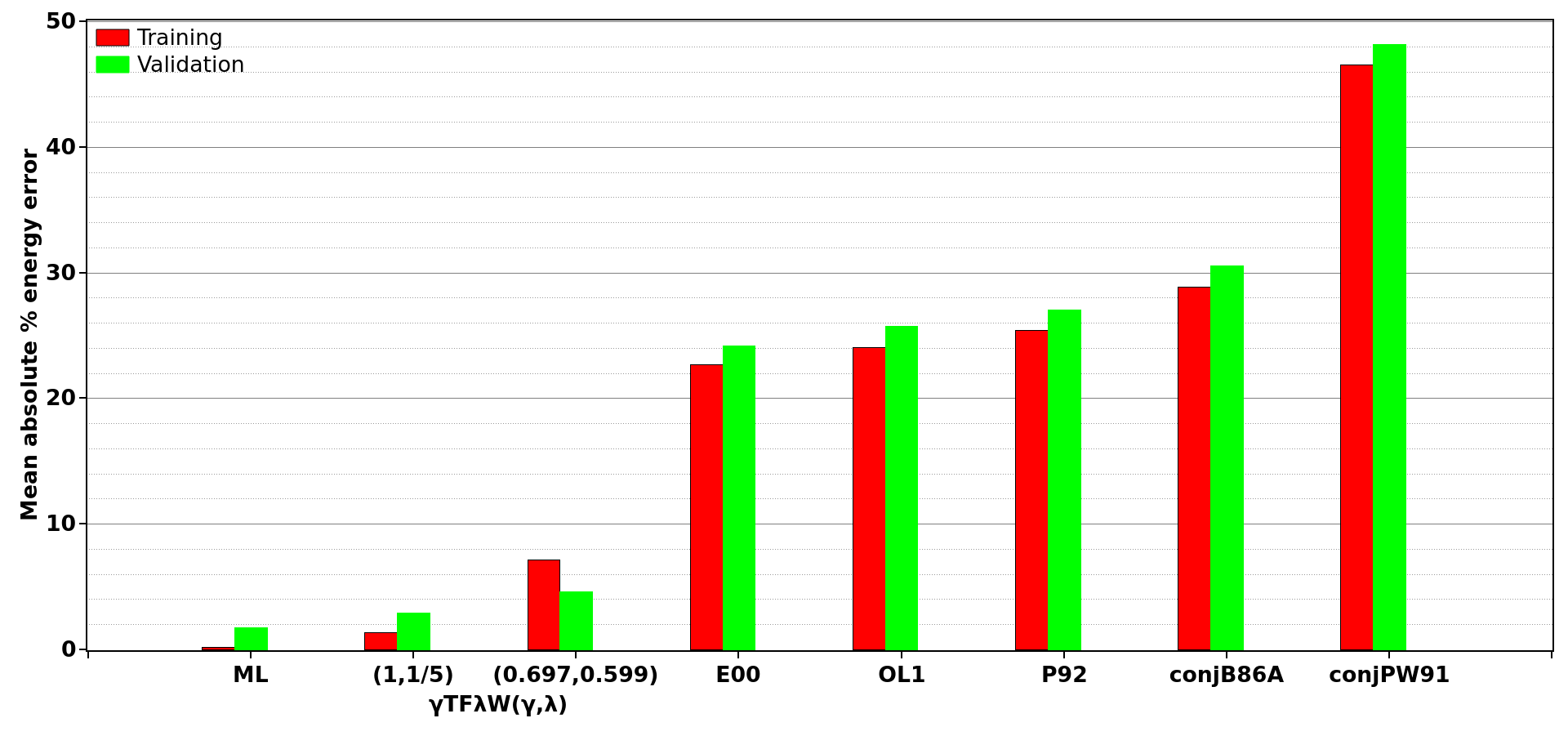}
\caption{Mean absolute percentage energy error for the training (red) and validation (green) sets.}
\label{fig: energy_error_all_functionals}
\end{subfigure}
\begin{subfigure}{1\linewidth}
\centering
\includegraphics[width=1\linewidth]{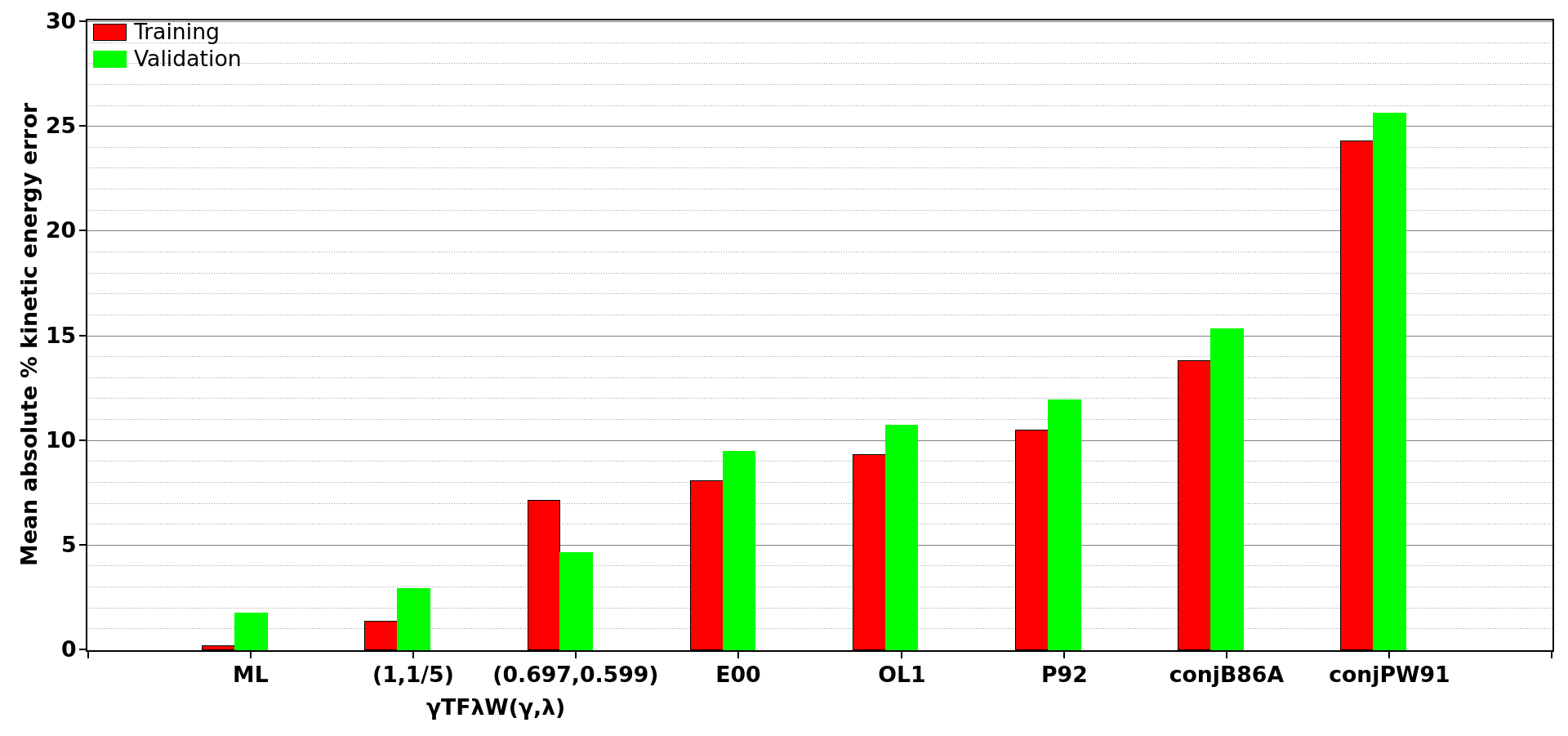}
\caption{Mean absolute percentage kinetic energy error for the training (red) and validation (green) sets.}
\label{fig: kinetic_energy_error_all_functionals}
\end{subfigure}
\caption{Mean absolute percentage (MAP) energy (a) and (b) kinetic energy errors for the training and validation sets for our ML-$\Omega$ OF the kinetic energy functional and 
a representative set of OF functionals, which is formed by the two of the $\gamma TF \lambda vW(\gamma,\lambda)$ family, the GGA functionals $E00, OL1, P92$, and the
conjoint functionals conjB86A and conjPW91.}
\label{fig: energy_and_kinetic_energy_errors_all_functionals}
\end{figure}

\begin{figure}
\centering
\includegraphics[width = 1\linewidth]{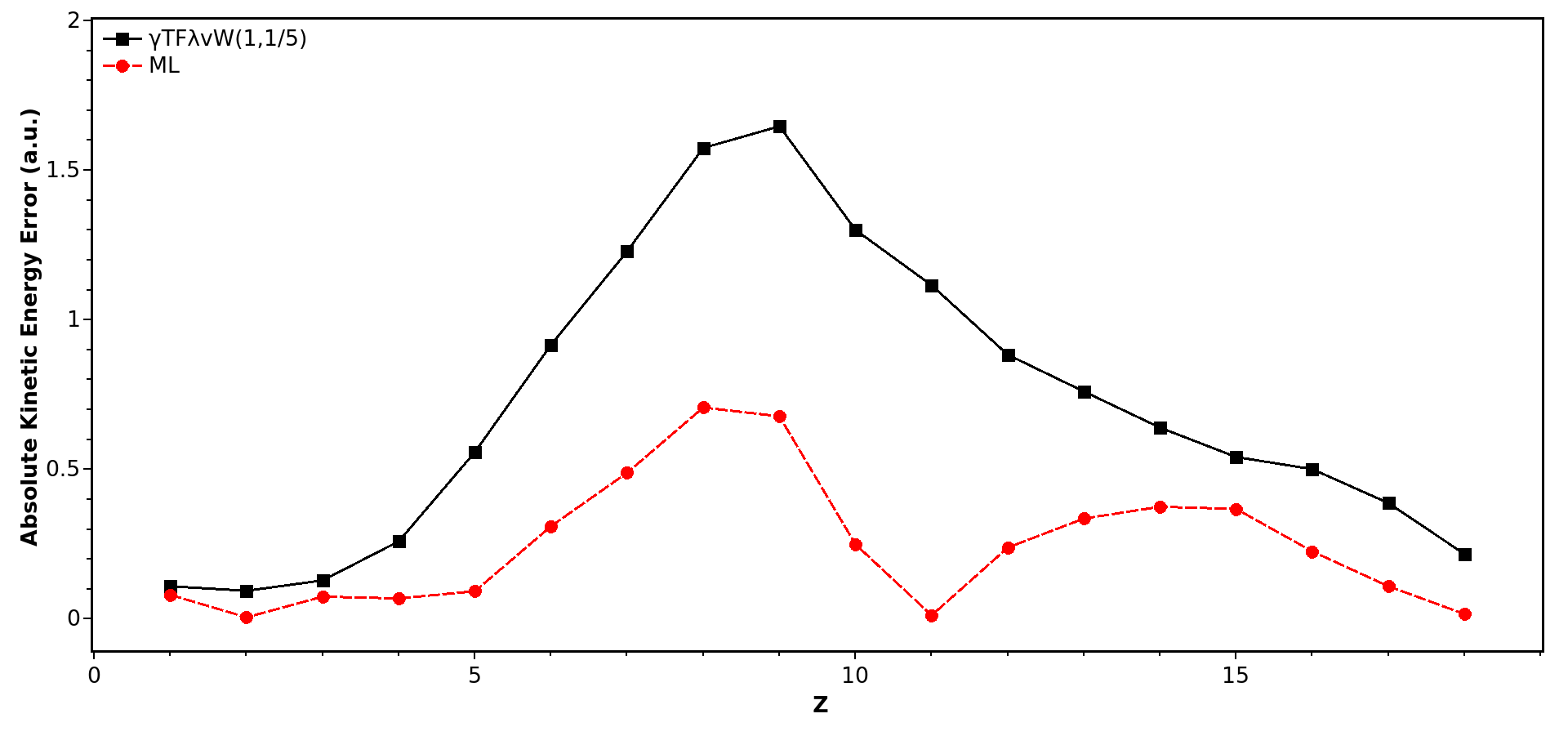}
\caption{Absolute error (AE) of the kinetic energy functionals ML-$\Omega$-OF functional (dashed red curve) and $\gamma TF\lambda W(1,1/5)$ model (solid black curve)
as a function of the atomic number (Z).}
\label{fig: KE_error}
\end{figure}

The minimization in Eq.(\ref{equ: Ts_OFDFT_ML}) must be performed considering the same restrictions as the TF case (Eqs. (\ref{equ: TF_restriction_normlaization} )-(\ref{equ: TF_restriction_positivity}) )\footnote{There is an additional restriction on $\phi$ :  $\int_{\mathfrak{R}^3} \frac{|\nabla \phi(\vec{r})|^2}{\phi(\vec{r})} d\vec{r} < \infty$. However, this restriction is automatically satisfied when $\phi$ is expanded in a basis set}. 
The ML-$\Omega$ model function Eq. (\ref{equ: Ts_OFDFT_ML}) can be trained with molecular data in the general case, the feature vector  $\vec{x}_i = (Z_{i1},\cdots, Z_{iM_{i}},\vec{R}_{i1},\cdots\vec{R}_{iM_{i}})$ contains the atomic numbers $(Z_{ij};j=1\,\cdots, M_i)$ and the corresponding nuclear positions $(\vec{R}_{ij};j=1,\cdots, M_{i})$. 
However, we only use here atomic data in the training set ( $\vec{x}_i = Z_i $) considering the energy of only five atoms He, Be, Ne, Mg and Ar as the target values. 
All KS-DFT calculations for the training  and validation sets were performed with the Amsterdam Density Functional (ADF-2019) software,
which is part of the suite AMS (see www.scm.com) \cite{ADF1998,ADF2001}.
The LDA Dirac's only exchange functional \cite{dirac1930note} was used for $E_{xc}$ along with a Slater quadrupole zeta plus 4 polarization 
functions (QZ4P) basis set \cite{VanLenthe2003}. All-electron calculations were performed for several spin configurations.
The lowest-energy spin configuration was used for the atomic training and validation data (see SI). All OF-DFT data was obtained by 
solving the OF-DFT electronic structure problem by direct minimization of the energy functional subject to  the proper restrictions 
(see benchmark calculations in reference \cite{vergara2023efficient} ) using the Dirac's exchange functional \cite{dirac1930note}.

\begin{figure}[ht]
\centering
\includegraphics[scale=0.25]{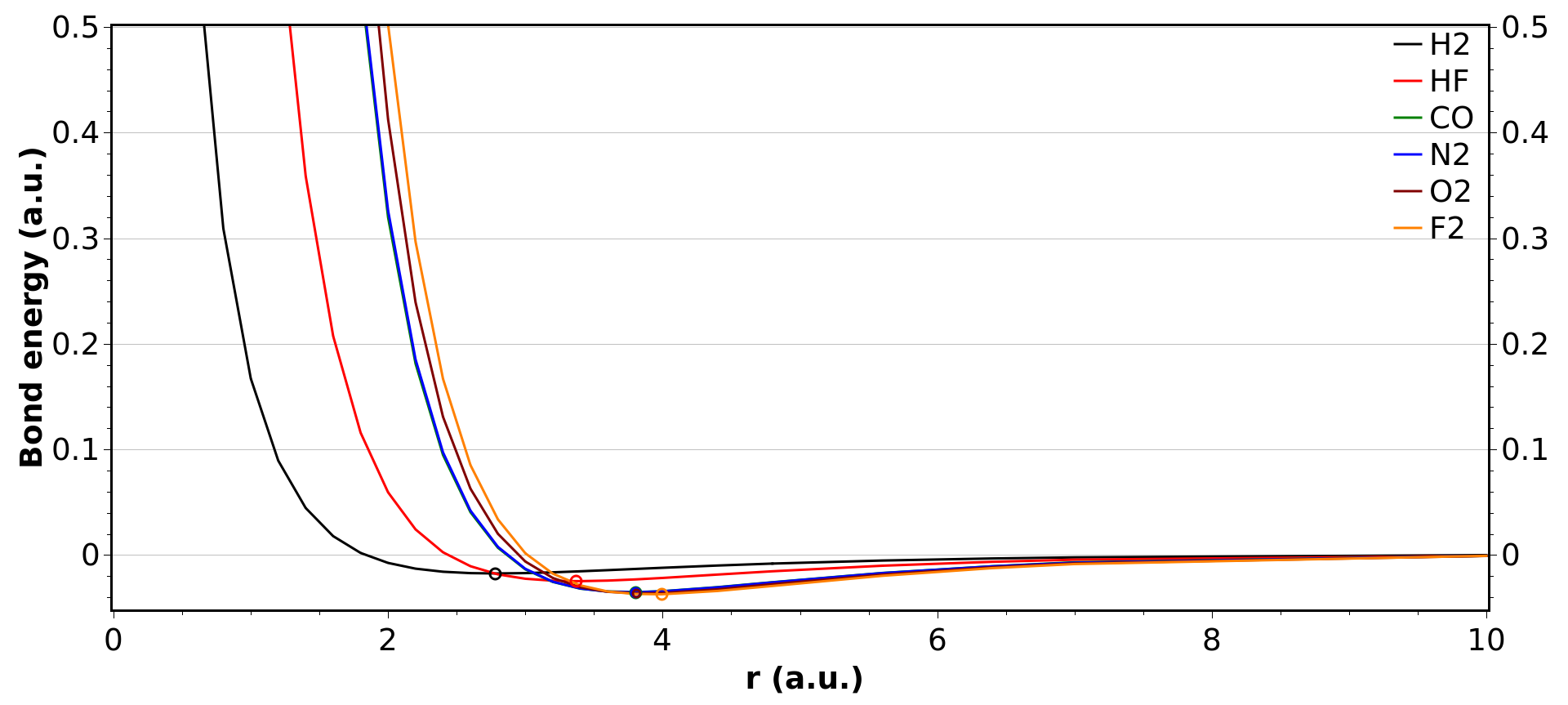}
\caption{Potential energy surface (PES) of  $H_2$,$HF$, $CO$,$N_2$,$O_2$, and $F_2$ molecules computed with our ML-$\Omega$-OF kinetic energy functional.}
\label{fig:PES_H2_ML_model}
\end{figure}

The weights after training for the ML-OF function can be seen in the last row of Table \ref{tab: w_values_S_and_TF}. 
The mean absolute percentage error (MAPE) for the training and validation sets, considering the latter as the energies of the atomic systems H to Ar (excluding the ones in the training set) are shown in Figure \ref{fig: energy_and_kinetic_energy_errors_all_functionals}. 
The lowest MAPE for both the training and validation sets for the kinetic and total energy corresponds to the ML-$\Omega$-OF functional.
It is important to mention that the total energies are computed by minimizing the corresponding OF-DFT functional (not just evaluating the functional) subject to the proper restrictions,
that is, by solving the OF-DFT electronic problem for each OF-DFT functional.
Thus, our ML-$\Omega$-OF functional, which could be also termed as $\gamma TF \lambda vW(0.964,1/4)$,
outperforms all the OF-DFT functionals of the representative group tested in this work \cite{karasiev2012issues,wesolowski2013recent} . 
This fact is of particular interest for the functionals $\gamma TF \lambda vW(1,1/5)$ and $\gamma TF \lambda vW(0.697,0.599)$ for which the ($\gamma$,$\lambda$) parameters
were pre-optimized \cite{yonei1965weizsacker,leal2015optimizing}. Fig. \ref{fig: KE_error} shows that the ML-$\Omega$-OF functional is closer to the reference values of $T_s$ than the best OF-DFT kinetic functional tested, the $\gamma TF \lambda vW(1,1/5)$ model, for all first 18 atoms (H to Ar) of the periodic table.
Interestingly, the information of only five atoms was used in the training set, showing the transferability (universality) of the ML-$\Omega$-OF functional.
To further evaluate the transferability/universality of the ML-$\Omega$-OF functional, it was tested on some diatomic molecules. Results can be seen in Table \ref{tab: ML_mol_energies_and_r_comparation}. 
The ML-$\Omega$-OF functional produces lower errors than the $\gamma TF \lambda vW(1,1/5)$ functional
when computing the bond distance and energies for all the diatomic molecules in Table \ref{tab: ML_mol_energies_and_r_comparation}.       
Although the ML-$\Omega$-OF functional 
is not fully competitive with KS-DFT accuracy, it improves the performance of any of the reported functionals belonging to the $\gamma TF \lambda vW$ model.
Moreover, the ML-$\Omega$-OF functional ($\gamma TF \lambda vW(0.964,1/4)$) does describe the correct bond breaking 
at the correct spin configuration for diatomic milecules of Table \ref{tab: ML_mol_energies_and_r_comparation} as can be seen in Figure \ref{fig:PES_H2_ML_model}, 
which remains an outstanding challenge for the LDA and some GGA functionals \cite{cohen2008insights}.

Notice that the ML-$\Omega$ method has some common features with the conventional regression methods for obtaining new prior-form functionals. 
The key difference of the ML-$\Omega$ model function, Eq. (\ref{equ: ML_model_function}), is that the proposed functional
is being globally minimized. This feature is very important in problems where the quantity of interest is the global minimum of a given functional,
as in the case of the ground state in Quantum Mechanics.  
For instance, Ryley et al. \cite{ryley2020robust} demonstrated the striking difference between the energies obtained at fixed density (from a reference KS-DFT calculation) 
and that obtained by solving the OF-DFT problem self-consistenly for a given OF-DFT functional.  
Such differences can be over 100 percent depending on the specific OF-DFT functional used (see Figure 5 in reference \cite{ryley2020robust}).                                                                                                                                       
In the training process of the ML-$\Omega$ function, the correct density is computed on the fly, which is the one that minimizes the 
ML-$\Omega$ functional (Eq. (\ref{equ: ML_model_function})).
This allows the optimization of the weights at the correct point of the ML-$\Omega$ functional,
at the correct value of $\phi$, that is, the optimum value.
Notice that in the ML-$\Omega$ method, the so-called density-driven functional error \cite{kim2013understanding,li2016understanding} is being minimized
while minimizing the loss function, Eq. (\ref{equ:Loss_function}). 
The ML-$\Omega$ method, however, retains the advantage of conventional functional regression by generating closed-form functionals,
whose functional derivative is easy to obtain analytically, which is not the case when using standard ML model functions like neural networks.
The functional derivative of a given OF-DFT functional is needed to solve the Euler-Lagrange OF-DFT equation for obtaining the system's electronic
structure. However, the high accuracy of the OF-DFT functionals obtained with the standard ML model functions is not inherited 
to their corresponding functional derivatives.\cite{snyder2012finding,snyder2015nonlinear,li2016understanding,meyer2020machine}  
It is worth noting that the method used in this work for solving the OF-DFT electronic structure problem does not need the functional derivative since it is
based on the zero-order evolutionary method recently introduced by us, which is explained in detail and benchmarked in reference \cite{vergara2023efficient}.

\begin{table}[]
\centering
\caption{Absolute error (AE) for the equilibrium distance (Err $r_b$) and dissociation energies (Err $E_b$); for the ML-$\Omega$-OF kinetic energy functional and 
the $\gamma TF\lambda W(1,1/5)$ \cite{yonei1965weizsacker,ludena2002reviews} functional with respect to the LDA values.}
\label{tab: ML_mol_energies_and_r_comparation}

\begin{tabular}{c|cc|cc}
 & \multicolumn{2}{c}{$\gamma TF \lambda W(1,1/5)$}  &  \multicolumn{2}{|c}{$ML-\Omega$} \\
      & Err $r_b$ & Err $E_b$ & Err $r_b$ & Err $E_b$     \\
\hline
\hline
$H_2$ & 1.53 & 0.12 & 1.30 & 0.11    \\
$HF$  & 1.94 & 0.20 & 1.59 & 0.19    \\
$N_2$ & 2.14 & 0.31 & 1.71 & 0.30    \\
$CO$  & 2.08 & 0.39 & 1.65 & 0.38    \\
$O_2$ & 2.01 & 0.21 & 1.50 & 0.20    \\
$F_2$ & 1.76 & 0.08 & 1.34 & 0.06    \\
      
\end{tabular}
\end{table}

In summary, we have introduced a machine learning model function ML-$\Omega$ that is inspired by the variational principle of physics.
When the method is provided the energies of hydrogen and two hydrogen-like atoms ($He^{+}$, $Li^{2+}$) for a given prior-form functional,
the ML-$\Omega$ method finds the correct Schrödinger functional and, from it, the fundamental Schrödinger equation.
Similarly, the ML-$\Omega$ approach finds the Thomas-Fermi energy functional within the DFT methodology.
The method is then applied to find the OF-DFT functional based on the $\gamma TF \lambda vW $ model.
The ML-$\Omega$-OF-DFT functional ($\gamma TF \lambda vW(0.964,1/4)$) outperforms all the OF-DFT functionals of
a representative group which includes two of the most accurate functionals of the $\gamma TF \lambda vW $ family.
The ML-$\Omega$-OF-DFT functional describes the potential energy surface (PES) properly in the stretched region of diatomic molecules.
Although the ML-$\Omega$-OF-DFT functional obtained in this work is not fully competitive with KS-DFT scheme in terms of accuracy, 
the current work provides a new path to generate novel highly-accurate OF-DFT functionals \cite{vergara2024}. 
The minimization of the the ML-$\Omega$ model function functional, Eq. (\ref{equ: ML_model_function}), permits the 
correlation of a target variable to an optimum value of such functional.
The Euler-Lagrange equation can be obtained after the training process allowing to go from data to differential equations which in turn
provides possible insight into the underlying physical, chemical, biological or engineering phenomena the data are derived from. This inherent interpretability is an important consideration in itself, and overcomes a common issue related with many modern ML approaches. 
The ML-$\Omega$ could also work with different functional forms of the model function.
However, this ingredient is not completely necessary since the evolutionary DE method used in the ML-$\Omega$
model function can be coupled with the techniques of symbolic regression \cite{angelis2023artificial},
an area of active focus for our group \cite{vergara2024b}. 
The flexibility of the ML-$\Omega$ model function, Eq. (\ref{equ: ML_model_function}), due to the infinite number of forms that the ML-$\Omega$ functional can have,
allows it to be applied to a wide range of phenomena in the natural sciences. 
In this paper, it was applied to three problems in Quantum Mechanics. 
In the first one, the ML-$\Omega$ successfully modeled one-electron systems, supposing that their experimental ground state energies were available. 
In the second problem, the energy functional for the homogenous electron gas, the TF-Dirac (TFD) functional, was obtained when 
the ML-$\Omega$ model function was trained with analytically-obtained energy values for the corresponding system.
Finally, in the third problem, the ML-$\Omega$ model function was trained with data of real n-electron systems (atoms)
for obtaining an OF expression for the independent electron functional $T_s$.
The proposed functional for the ML-$\Omega$ model function was based on the $\gamma TF \lambda vW $ functional,
which is known to be a upper bound for $T_s$. \cite{dreizler2012density}  
The restrictions on the function $\phi$ were aimed to produce physically acceptable solutions in each problem. 
  
Since the ML-$\Omega$ model function allows one to go directly from training data to a differential equation, which itself potentially describes the underlying
physics, chemistry, or biology phenomena, the ML-$\Omega$ method can be considered at the intersection of ML and the natural sciences.

\section*{Acknowledgment}
The authors thanks Dr. David C. Thompson for proofreading 
the manuscript and for insightful comments.
The authors also acknowledge the "Secretaría de Investigación y Posgrado", 
Instituto Politécnico Nacional-México,
for financial support (project SIP-20240537).
U.A.V.B. thanks CONACYT-Mexico for the doctorate fellowship.

\bibliographystyle{unsrt}
\bibliography{biblio}

\end{document}